\documentclass[fleqn,usenatbib]{mnras}

\usepackage{newtxtext,newtxmath}

\usepackage{graphicx}	
\usepackage{subcaption}
\usepackage[T1]{fontenc}
\usepackage{ae,aecompl}
\usepackage{tabularx}

\usepackage{amsmath}	
\usepackage{amssymb}	
\usepackage{natbib, booktabs, multirow, tabularx} 
\usepackage{hhline}

\usepackage{float}
\restylefloat{figure}

\title[FRB widths]{Interpreting the distributions of FRB observables}

\author[Liam Connor]{
Liam Connor,$^{1}$
\\
$^{1}$Anton Pannekoek Institute for Astronomy, University of Amsterdam, Science Park 904, 1098 XH Amsterdam, The Netherlands\\
}

\date{Accepted XXX. Returned YYY; in original form ZZZ}

\pubyear{2019}

\begin{document}
\label{firstpage}
\pagerange{\pageref{firstpage}--\pageref{lastpage}}
\maketitle

\begin{abstract}
Fast radio bursts (FRBs) are short-duration radio transients of unknown origin. Thus far, they have been blindly detected at millisecond timescales with dispersion measures (DMs) between 110--2600\,pc\,cm$^{-3}$. However, the observed pulse width, DM, and even brightness distributions depend strongly on the time and frequency resolution of the detection instrument. Spectral and temporal resolution also significantly affect FRB detection rates, similar to beam size and system-equivalent flux density (SEFD). I discuss the interplay between underlying FRB properties and instrumental response, and provide a generic formalism for calculating the \textit{observed} distributions of parameters given an intrinsic FRB distribution, focusing on pulse width and DM. 
I argue that if there exist many FRBs of duration $<<$\,1\,ms (as with giant pulses from 
Galactic pulsars) or events with high DM, they are being missed due to the deleterious effects of smearing. I outline how to optimise spectral and temporal resolution for FRB surveys that are throughput-limited. I also investigate how such effects may have been imprinted on the distributions of FRBs at real telescopes, like the different observed DMs at ASKAP and Parkes. Finally, I discuss the impact of intrinsic correlations between FRB parameters on detection statistics. 
\end{abstract}




\section{Introduction}

Fast radio bursts (FRBs) are a new class of extragalactic radio 
transient whose origins remain a mystery. The dispersion measures (DM) 
of observed FRBs have been between 110--2600\,pc\,cm$^{-3}$,
significantly exceeding the expected 
contribution from the Milky Way along their 
respective lines-of-site. Roughly 70 sources 
of FRBs have been found to date \citep{petrofffrbcat}. 
They have now been discovered at six different 
telescopes around the world, in bands between 400\,MHz and 8\,GHz
\citep{lorimer07, thornton-2013, spitler2014, masui-2015b, caleb2017, shannon2018, chime2019a, Gajjar2018}. 
Of the known FRB-emitting sources, two \citep{spitler2016, chime2019r2} 
have been found to repeat; all others have been once-off events.
Blindly-detected events have had duration 0.64--21\,ms, 
but thanks to coherent dedispersion of dumped raw 
voltage data \citep{michilli2018,farah2018} and 
pulse fitting \citep{ravi2016,chime2019a}, some FRBs are known to 
have structure on microsecond timescales. 

Several broad-band, wide-field 
surveys are expected to be transformative for the FRB field, 
and are presently either operational or in the late commissioning stage.
These include the Canadian Hydrogen Intensity Mapping 
Experiment (CHIME; \citet{chime2018overview}), 
the Australian Square Kilometre Array Pathfinder (ASKAP; \citet{bannister2017}), 
UTMOST \citep{utmost2017}, the APERture Tile in Focus (Apertif; \citet{leeu14}), 
and Meer TRAnsients and Pulsars (MeerTRAP) \citep{sanidas2018}.
As the field transitions from small-number statistics 
to hundreds or thousands of new FRBs per year, our 
statistical tools must mature accordingly. This will 
include understanding the interplay 
between the underlying source distribution, our
instruments, and FRB detection statistics. 
The underlying source distribution 
refers to intrinsic physical properties of the FRB 
population, including 
energetics, redshift distribution, and spectral behaviour. 
Detection statistics refers to FRB observables, 
such as the observed brightness distribution and event detection rates. 
The mapping between intrinsic and observed properties is central to this paper.

Several groups have considered brightness completeness effects in 
order to accurately calculate event rates \citep{keane2015, connor2016b}.
Others have studied the effects 
of the redshift, luminosity, and spectral index distributions of FRBs on 
their detected source counts \citep{Oppermann16, caleb2016a, vedantham2016, connor2017, niino2018}. 
In \citet{macquart2018_2}, the authors derive many of 
the relations between such physical properties 
from first principles for cosmological FRBs.

Less attention has been paid to the subtle ways in which instrumental 
response can affect the distributions of FRB observables, or how 
different temporal and spectral resolution can alter by orders-of-magnitude
the detection rate at two otherwise-identical telescopes. 
These selection effects are particularly 
important for interpreting the observed DM and brightness statistics 
of FRBs, which contain information about their 
spatial and luminosity distributions. The new era of sensitive, wide-field surveys 
will require a low-level understanding of how to tease out the underlying 
DM distributions and source counts from what is observed. 
Beyond DMs, luminosities, and redshifts, such understanding will 
also enable us to constrain the intrinsic spectrum of FRBs, 
volumetric event rates, energy cut-offs, and intrinsic correlations 
between FRB variables.
It will also 
affect various proposed FRB applications. 
For example, missing highly-dispersed events due to instrumental selection effects
may preclude the use of FRBs in cosmology \citep{mcquinn2014, masui2015a, Madhavacheril2019}. 

In Sect.~\ref{sect-formalism} of this paper I develop a generic formalism for calculating 
detection rates at a given instrument, based on the 
underlying FRB distribution.
This includes a method for determining the 
\textit{observed} distribution of an FRB parameter given its
\textit{true} distribution, where I focus here on pulse width 
and DM. I then discuss optimising time and 
frequency resolution 
on an instrument once something is known about the 
true distributions of DM, pulse width, and source counts. 
In Sect.~\ref{sect-data} I apply these tools to real data and investigate 
how such effects might have been imprinted on the 
detection statistics of FRBs at telescopes such as ASKAP and 
Parkes. Finally, in Sect.~\ref{sec-discussion} I investigate 
how correlations between FRB parameters can alter the 
relationship between the observed and true distribution of FRBs. 

\section{Detection statistics}
\label{sect-formalism}

The detection rate, $\mathcal{R}$, of a blind single-pulse survey is given 
by an integral of differential event rate 
over all relevant parameters, $\mathbf{\lambda} = [{\lambda_1,\ldots,\lambda_m}]$, in 
the region of phase space, $\mathbf{\Lambda}$, 
to which that survey is sensitive,


\begin{equation}
    \mathcal{R} = \int \cdots \int_\mathbf{\Lambda}\, n(\lambda_1,\lambda_2,\ldots,\lambda_m) \,d\lambda_1 \!\cdots d\lambda_m.
    \label{eq-1}
\end{equation}

\noindent Here $n(\lambda_j)$ is the number density of 
events around parameter $\lambda_j$. Formally, this is a 
differential intrinsic event rate,

\begin{equation}
    n(\lambda_j) \equiv \frac{\mathrm{d\,N}}{\mathrm{d}\lambda_j}.
    \label{eq-diff}
\end{equation}

\noindent For example, $n(s)$ would be the differential brightness distribution 
of FRBs before any selection effects. As I show in Sect.~\ref{sec-discussion}, 
correlations between the different distributions $n(\lambda_j)$ can 
affect detection statistics.

Typically, FRB detection rates 
have been computed using only beam size and a brightness threshold, 
where the integral is calculated over location on the sky, denoted by the unit vector, $\hat{k}$,
and a measure of pulse strength, $s$, whether flux density, 
fluence, or signal-to-noise ratio (S/N). The equation,

\begin{align}
    \mathcal{R} &= \int\displaylimits_{4\pi} \int\displaylimits_{s_{\mathrm{m}}} n(\hat{k}, s)\,
    \mathrm{d}\hat{k}\mathrm{d}s
    \label{eq-rate1}
\end{align}

\noindent is often simplified with the
assumptions that the radio beam is a two-dimensional top hat, 
the events are isotropic within a beam, and the 
brightness distribution is a power-law such that, 
$n(s)\propto s^{-\alpha-1}$. 
This results in the familiar equation,

\begin{align}
    \mathcal{R} &\propto \Omega\,s_{\mathrm{m}}^{-\alpha},
    \label{eq-ratesimple}
\end{align}
\vspace{3pt}

\noindent where $\Omega$ is a measure of the telescope's 
field of view (FoV),
$s_{\mathrm{m}}$ is a minimum detectable brightness, and $\alpha$ is the
source counts power-law index, such that 

\begin{equation}
\alpha\equiv\frac{\partial{\log n(>s)}}{\partial{\log s}}.
\label{eq-4}
\end{equation}
\vspace{3pt}

Several authors have highlighted the importance of 
integrating over the full beam, due to the dependence of 
$s_{\mathrm{m}}$ on where in that beam the FRB arrives
\citep{vedantham2016, Lawrence2017, macquart2018_1}. By 
considering the full beam,

\begin{equation}
    \mathcal{R} = \int\displaylimits_{4\pi} n(\hat{k})\!\!\int\displaylimits_{s_{\mathrm{m}}(\hat{k})} s^{-\alpha-1}\,
   \mathrm{d}s\,\mathrm{d}\hat{k},
    \label{eq-rate2}
\end{equation}

\noindent one can account for the detection of bright, 
but rare, events in the side-lobes, whose sky coverage
are much larger than 
the primary beam. This is valuable for estimating FRB sky rates 
and for comparing detection rates between surveys with 
different beam shapes. 
The resultant brightness distribution shape is not 
affected by the beam response so long as the underlying 
source counts distribution is a power-law and the dynamic range 
between the brightest and dimmest event is large. If one of those 
conditions is not met, the observed brightness distribution's 
logarithmic slope can deviate from that of the input, 
and its functional form can stray from power-law.

However, Eq.~\ref{eq-ratesimple} 
not only neglects beam effects, but  
ignores the distributions of several other important parameters 
and the way they differentially affect 
surveys with disparate properties.
If I also include 
DM and intrinsic pulse width, $t_i$, on top of sky location
and brightness, I have $\mathbf{\lambda} = [\hat{k}, s, \mathrm{DM}, t_i]$ and the 
integral in Eq.~\ref{eq-1} becomes,


\vspace{7pt}
\begin{align}
     \mathcal{R} &= \int n(\hat{k}, s, \mathrm{DM}, t_i)\, \mathrm{d}\hat{k}\,\mathrm{d}s\,\mathrm{dDM}\,\mathrm{d}t_i
     \\
     &= \int\displaylimits_{4\pi} n(\hat{k}) \int\displaylimits_{0}^\infty n(t_i) \!\int\displaylimits_{\mathrm{DM_{min}}}^{\mathrm{DM_{max}}} \!
    n(\mathrm{DM})\, \int\displaylimits_{s'_{\mathrm{m}}}^\infty n(s)\,\mathrm{d}S\,\mathrm{dDM}\,\mathrm{d}t_i\,\mathrm{d}\hat{k}.
\label{eqn-integral}
\end{align}
\vspace{7pt}

\noindent This equation calculates the number of detectable events 
per unit time for a survey with $\mathrm{DM_{min}}$, $\mathrm{DM_{max}}$, 
given an underlying FRB distribution 
described by $n(t_i)$, $n(\mathrm{DM})$, and $n(s)$. Crucially, 
the minimum detectable single-pulse brightness, $s'_{\rm{m}}$, 
depends on all of the former inputs. It will depend on 
where in the beam the FRB arrives; it is also increased 
by the deleterious effects of temporal smearing due to intra-channel 
dispersion and the instrument's finite time sampling. 

These effects all
reduce detection rate
and are not necessarily 
small corrections:
If the functional form of $n(t_i)$ were similar 
to $n(s)$ (source counts), then halving one's instrumental 
smearing would have as much impact as doubling 
the telescope's collecting area or cutting in half 
its system temperature with a cooled receiver. 

In this 
paper I consider two simple intrinsic DM distributions. The first 
traces the star-formation rate, which is a top-heavy distribution 
with many highly-dispersed events. I follow \citet{niino2018}
in using a DM redshift relation give by,

\begin{equation}
    \mathrm{DM}(z) = 1000\,z \,\,\,\mathrm{pc\,cm}^{-3}.
\end{equation}
\vspace{3pt}

\noindent Combining this with an FRB number density 
in redshift, $\rho(z)$, and the comoving volume 
element per redshift bin, $\frac{dV}{dz}$, the 
number of FRBs per DM is,

\vspace{3pt}
\begin{equation}
    \frac{dn}{d\mathrm{DM}} = \frac{\rho(z)}{(1+z)} \frac{dV}{dz} \frac{dz}{d\mathrm{DM}},
\end{equation}
\vspace{3pt}

\noindent where I use, 
\vspace{3pt}

\begin{equation}
    \rho(z) = \rho_{SFR}(z) \propto \frac{(1+z)^{2.5}}{1+((1+z)/2.9)^{5.6}}
\end{equation}
\vspace{3pt}

\noindent from \citet{madau2014}.
The second is a Gaussian 
distribution peaking at 1000\,pc\,cm$^{-3}$ with standard 
deviation 500\,pc\,cm$^{-3}$, corresponding to a scenario 
where FRBs are primarily dispersed local to the source. In 
reality, the intrinsic DM distribution of FRBs will depend 
on their luminosity function \citep{niino2018, macquart2018_1}. But by using
both a top- and bottom-heavy DM distribution, most plausible outcomes are spanned. 
In the following subsections, I consider the implications of Eq.~\ref{eqn-integral} and the aforementioned DM 
distributions on the observed properties of 
FRBs in the presence of real instrumental effects.

\subsection{Temporal smearing}
The observed pulse width is a quadrature sum of the 
intrinsic pulse width, the dispersion-smearing timescale, and
the sampling time \citep{cordes2003},

\begin{equation}
    t_{\mathrm{obs}} = \sqrt{\,t_i^2 + t^2_{\mathrm{DM}} + t^2_s}.
    \label{eq-tobs}
\end{equation}

\vspace{5pt}

\noindent I define the intrinsic width, $t_i$, not as the emitted pulse width, but as the 
duration of the pulse when it arrives at a radio telescope, before 
instrumental smearing. 

\begin{equation}
    t_i = \sqrt{\,t^2_e(1+z)^2 + \tau^2}.
\end{equation}
\vspace{5pt}

\noindent Here, the FRB from redshift $z$ had a rest-frame emission 
width at the source, $t_e$, and has a final
scattering timescale $\tau$.

I do not attempt to model the temporal 
scattering, emission width, or FRB redshift distributions, 
and instead only include the understood phenomenon of instrumental 
pulse broadening. I ignore the discretisation of pulse widths 
introduced by single-pulse search software \citep{keane2015}, which usually 
choose a finite set of boxcar widths. I have also 
assumed that intra-channel dispersion smearing produces a Gaussian 
profile, even though it is closer to the intrinsic profile convolved with a box-car. An upcoming 
FRB detection software challenge will offer more realistic insights into 
such single-pulse search effects
\footnote{https://www.eyrabenchmark.net/benchmark/e515e4da-8044-45ad-8464-317c445a1bd8}.

Finite frequency resolution 
leads to a dispersion delay within individual 
frequency channels, which artificially broadens 
a dispersed pulse. This intra-channel dispersion smearing 
is linear in DM as,

\begin{equation}
    t_{\mathrm{DM}} = \rm{A\,DM},
\end{equation}

\vspace{5pt}

\noindent where the proportionality constant is

\begin{equation}
    \mathrm{A} = 8.3 \left( \frac{\Delta\nu}{1\,\mathrm{MHz}} \right )
    \left( \frac{\nu_c}{1\,\mathrm{GHz}} \right )^{-3} \, \, \mu \mathrm{s},
\end{equation}

\vspace{5pt}

\noindent and depends on 
the instrument's frequency resolution, $\Delta\nu$,
and central observing frequency, $\nu_c$. The 
relevant timescales in this paper are listed in Table.~\ref{tab-timescales}.




The minimum brightness detection threshold in the 
absence of broadening is denoted by, $s_{m,o}$. This 
threshold will be 
transformed by the beam response 
and the smearing terms in the following way,

\begin{equation}
    s'_{\mathrm{m}} = s_{\mathrm{m},o} \,\, b(\hat{k})^{-1} \, \left ( \frac{t_i^2 + t_s^2 + \mathrm{A}^2\,\mathrm{DM}^2}{t_i^2} \right )^{1/2} ,
    \label{eq-3}
\end{equation}

\vspace{5pt}

\noindent where the beam response, $b(\hat{k})$, is 
assumed to be unity on-axis. 

Ignoring beam effects for now, one can investigate how 
event detection rates depend on the sensitivity 
reduction from temporal smearing 
(for beam effects, see, e.g., \citep{ravi2019, james2019b}). 
Taking $\mathcal{R}_{\hat{k}}$ to be the 
detection rate per unit solid angle and evaluating 
the brightness component of the integral in Equation~\ref{eq-3} 
with a power-law $n(s) = s^{-\alpha-1}$, 
one gets, 

\begin{table}
\caption{Relevant FRB width timescales}

\begin{center}
\begin{tabular}{llllll}
            &                               &  &  &  &  \\ \cline{1-4}
            
$t_e$                & restframe emitted width       &  &  &  &  \\
$\tau$   & scattering timescale          &  &  &  &  \\
$t_i$                & intrinsic width (at receiver) &  &  &  &  \\
$t_{\mathrm DM}$           & intra-channel dispersion-smearing timescale         &  &  &  &  \\
$t_s$                & sampling time                 &  &  &  &  \\
$t_{\mathrm obs}$          & final observed pulse width    &  &  &  & 
\end{tabular}
\label{tab-timescales}
\end{center}
\end{table}


\begin{equation}
    \mathcal{R}_{\hat{k}} = \frac{s_{\mathrm{m},o}^{-\alpha}}{\alpha}\int\displaylimits_{0}^\infty \!n(t_i)\!\! \int\displaylimits_{\mathrm{DM_{min}}}^{\mathrm{DM_{max}}} \!\!\!
    n(\mathrm{DM}) \left ( \frac{t_i^2 + t_s^2 + \mathrm{A}^2\,\mathrm{DM}^2}{t_i^2} \right)^{-\alpha/2} \!\!\!\!\!\!\mathrm{dDM}\,\mathrm{d}t_i .
    \label{eq-rate_sa}
\end{equation}

\noindent Because $s'_m \geq s_{m,o}$ and 
$\alpha$ is positive, the detection rate
will always be decreased by the finite time and frequency resolutions 
of real telescopes. The degree to which FRBs are ``missed''
due to smearing will depend on their underlying brightness, DM, and width 
distributions. The latter two have been neglected in the FRB statistics
literature, but are very important if there exist considerable numbers 
of narrow and high-DM FRBs. 

Qualitatively, 
if $\alpha$ is large and the brightness distribution is 
steep, a greater fraction of events will be missed 
due to smearing because there are many events
close to the detection threshold that will 
fall below it. If the distribution 
of intrinsic pulse widths, $n(t_i)$, is such that 
there are lots of events narrower than the sampling 
or the dispersion-smearing timescale, surveys with poor 
time and frequency resolution will take a large hit 
in their event rates. The same is true 
if the underlying DM distribution of FRBs is weighted towards 
highly dispersed events. However, it is not just 
the overall detection rates that will be affected:
The observed DM and brightness distributions, as well 
as the inter- and intra-survey correlations between 
those quantities, will also be changed. 

\begin{figure}
  \centering
    \includegraphics[width=0.47\textwidth]{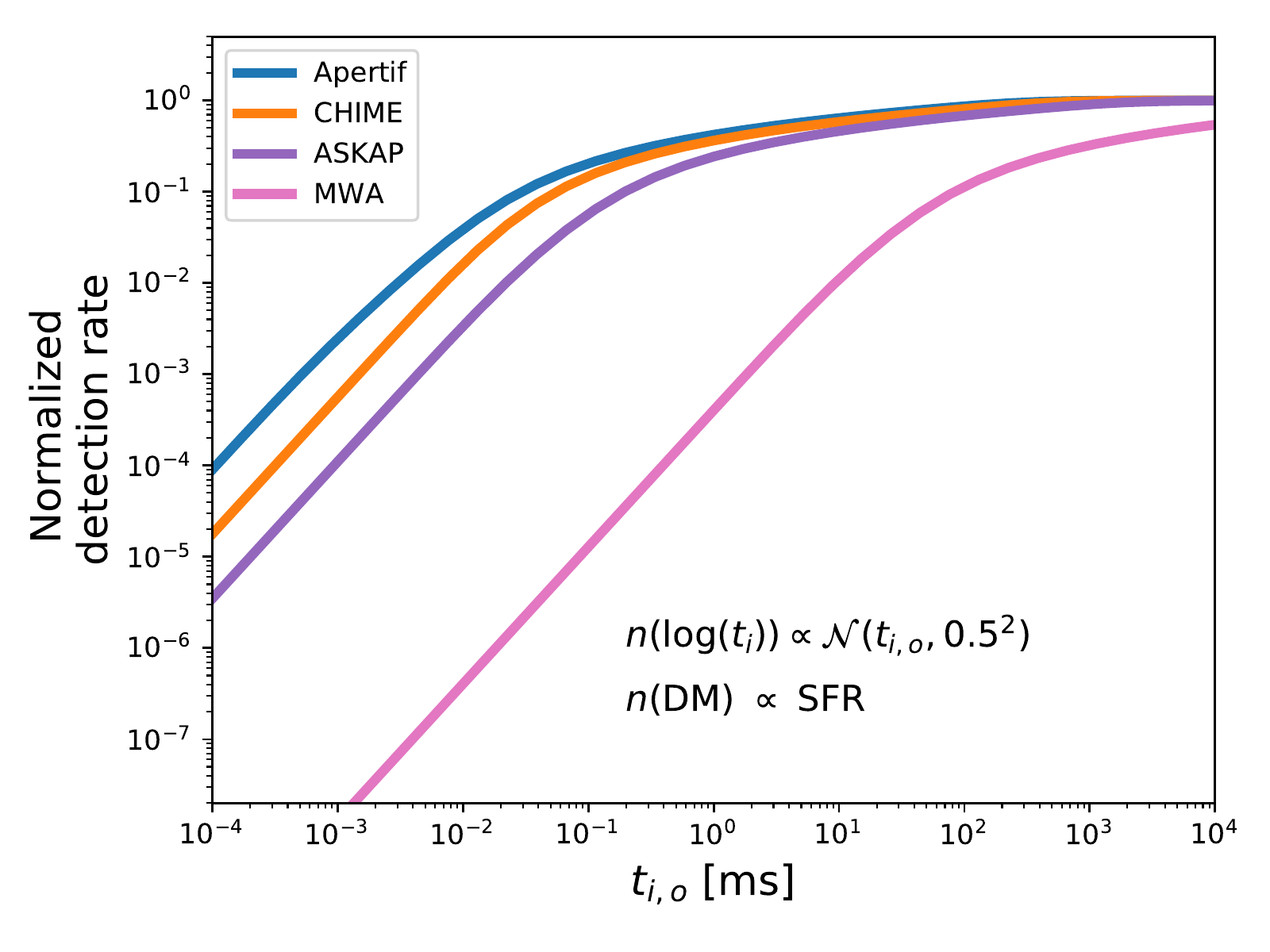}
  \vspace{-2pt}
  \caption{The FRB detection rate at several surveys normalised 
  to their rate if they had perfect time and frequency 
  resolution. This value is plotted as a function of 
  mean pulse width, $t_{i,o}$, of a lognormal intrinsic 
  width distribution with standard deviation $\log0.5$. 
  The DM distribution follows the star-formation rate.}
  \label{fig-rateti}
\end{figure}

The 2D integral in Eq.~\ref{eq-rate_sa} is calculated 
for Fig.~\ref{fig-rateti} using a DM distribution that follows the star-formation rate (SFR) and a lognormal intrinsic pulse width distributions with varying means. Clearly the FRB detection rate of an 
instrument depends strongly on the underlying pulse widths, 
drastically so in the case of low-resolution back-ends. 
This must be accounted for 
when rates between surveys are compared, or when detection 
rates are forecasted. The survey parameters assumed in this paper are 
listed in Table.~\ref{tab-parms}.

\subsection{Observed DM distribution}
\label{sec-dmdist}
The combination of temporal smearing and the 
unknown, intrinsic pulse width and DM distributions 
will differentially affect detection rates at 
each FRB survey. They will also alter the 
shape of distributions of various observables, including
DM. One can calculate the fraction of recovered events, $\eta$,  
at each DM, given a brightness distribution,
intrinsic pulse width distribution, and instrumental 
parameters. Starting from Equation~\ref{eq-rate_sa}, 
this is,

\begin{align}
    \eta(\mathrm{DM}) &\equiv \frac{n_{\mathrm{obs}}\left(\mathrm{DM}\right)}{n\,(\mathrm{DM})}
    \\
        &= \frac{1}{C} \int\displaylimits_{0}^\infty \!n(t_i)\, \left ( \frac{t_i^2 + t_s^2 + \mathrm{A}^2\,\mathrm{DM}^2}{t_i^2} \right)^{-\alpha/2} \mathrm{d}t_i
\end{align}

\noindent where $C$ is a normalising constant representing 
the number of events that would have been detected with 
perfect time and frequency resolution,

\begin{equation}
    C \equiv \int\displaylimits_{0}^\infty \!n(t_i)\, \mathrm{d}t_i.
\end{equation}

\begin{figure}
    \centering
	\includegraphics[width=0.46\textwidth,trim=0cm 0cm 0cm 0cm]{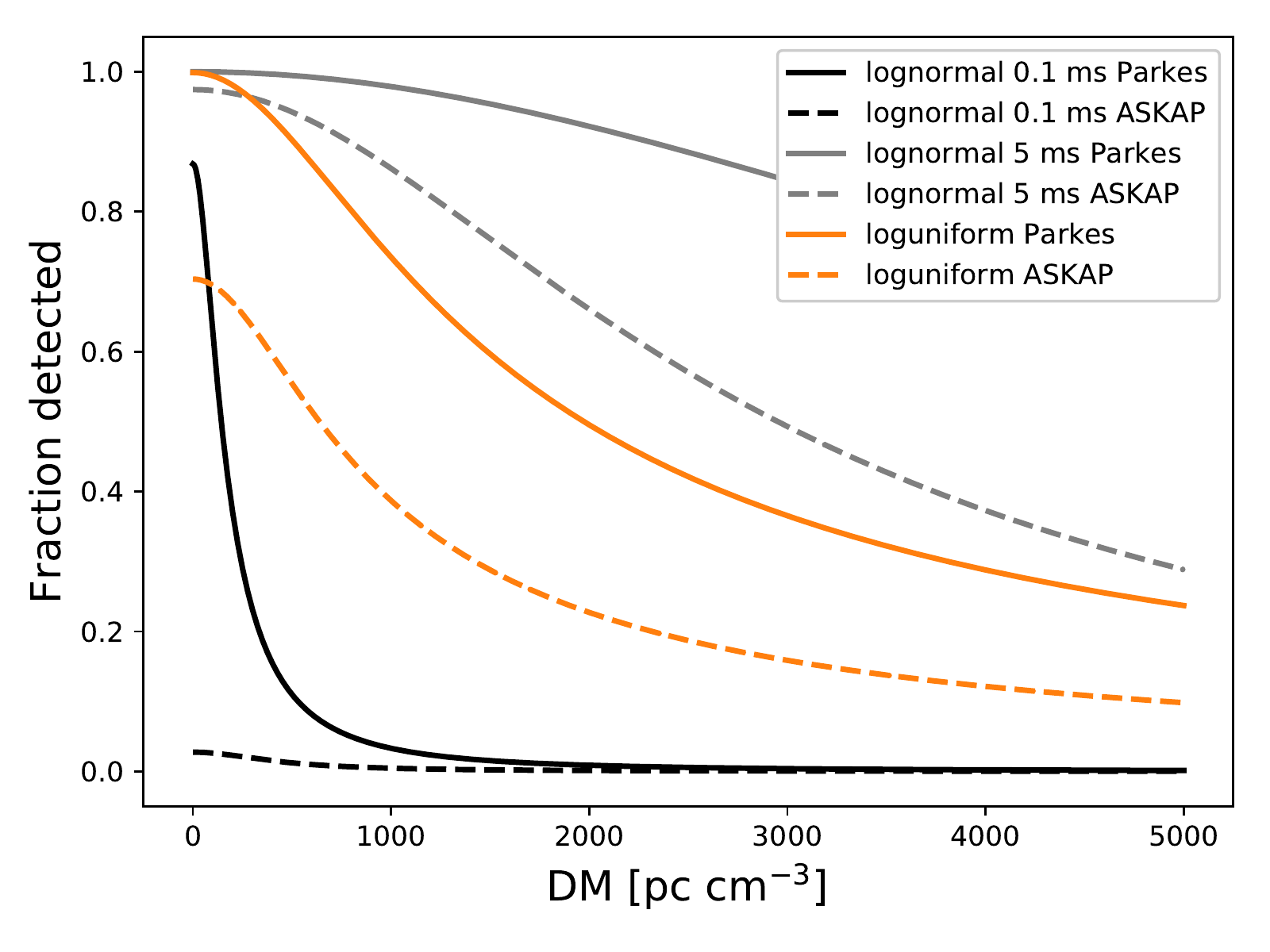}
    \caption{The fraction of events detected by Parkes (solid) and ASKAP 
    (dashed) compared to the same survey without smearing, plotted 
    as a function of DM for three different intrinsic pulse 
    width distributions. The log-normal curves have standard deviation $0.25$. 
    Without knowing the 
    intrinsic width distribution, one cannot
    know the transfer function between true and 
    observed DMs.}
    \label{fig-recovered}
\end{figure}

In Fig.~\ref{fig-recovered} I show how the distributions of 
FRB parameters intermix with instrumental effects and can 
significantly affect both detection rates and observed DM distributions.  
The fraction of recovered FRBs, $\eta$, is plotted as a function 
of DM for Parkes and ASKAP, using three different intrinsic 
pulse width distributions: lognormal, peaked at 100\,$\mu$s;
lognormal, peaked at 5\,ms; and loguniform. The lognormal integrals 
are calculated over four order-of-magnitude centered on the mean. 
The loguniform distribution is summed from 0.01--100\,ms. 

Clearly, a large fraction of FRBs are missed at 
DM\,$\gtrapprox$\,1000\,pc\,cm$^{-3}$ for many width 
distributions, especially using ASKAP's current 
temporal and spectral resolution. It should not be 
surprising that all $\sim$\,70 
FRBs have had DM\,$\leq$\,2600\,pc\,cm$^{-3}$: Either 
such high-DM events truly are rare or they are common 
and smeared below our detection thresholds. 
And given the intrinsic pulse widths
of FRBs need not be that narrow for current back-ends to miss 
them, it seems likely that many high-DM events are indeed being lost.
Highly-dispersed bursts probe either the most dense 
FRB environments or the highest redshifts, so they will be of 
great scientific interest.  
Therefore, surveys with the ability to 
trade time resolution for higher frequency resolution 
could alter their DM response function, making 
their instrument more sensitive to high-DM events. I discuss 
such trade-offs in Sect.~\ref{sect-optimise}.

If the DM distribution of FRBs 
is top-heavy (i.e. lots of highly dispersed events), then 
the detection rates of all current surveys are 
considerably suboptimal. Perhaps more importantly, 
$\eta$ represents a transfer function between the 
true and observed DM distribution, and its shape depends 
strongly on the instrument's time and frequency resolution. In 
Sect.~\ref{sect-data} I investigate whether this could produce 
the DM/fluence relationship found between ASKAP and Parkes \citep{shannon2018}. 

\begin{table}
\begin{center}
\caption{Survey parameters used in this paper. UTMOST 
increased its time/frequency resolution in 2018 after its 
first three detections.}
\begin{tabular}{cccccc}
\textbf{Survey}  & $t_s$ [ms] & $\Delta \nu$ [MHz] & $K$ [ms MHz] & $\nu_c$ [GHz] &  \\ \cline{1-5}
ASKAP   & 1.3    & 1.0   & 1.3 & 1.4 &  \\
Parkes  & 0.064 & 0.39   & 0.19   & 1.4 &  \\
Apertif & 0.041 & 0.195  &    0.08 & 1.4 &  \\
CHIME   & 0.983    & 0.024 &   0.024  & 0.6 &  \\
UTMOST2017$^\dagger$  & 0.655     & 0.780    &   0.511 & 0.82 & \\
UTMOST2018$^*$  & 0.327     & 0.097    &   0.032  & 0.82 & \\
MWA   &  500  & 1.24 &  620  & 0.185 &  \\
\end{tabular}
\label{tab-parms}
\end{center}
\end{table}

\subsection{Observed width distribution}
\label{sec-widthdist}

With pulse duration, 
calculating the observed distribution is less trivial than for DMs 
because widths are transformed at the instrument by smearing. 
In contrast, the observed DM of a given 
FRB is the same at the DM with which is arrives 
at our receivers. The distribution of the 
smeared widths is a probability density function (PDF) 
of a variable that has undergone a non-linear transformation. 
Nonetheless, in many cases 
the new distribution can be solved analytically
if the transformation is known, which it is 
for smearing.

If the PDF of a real random variable is known, 
the distribution of a function, $f$, of that variable 
can be calculated, as long as $f$ is invertible\footnote{https://www.stat.washington.edu/$\sim$nehemyl/files/UW\_MATH-STAT395\_functions-random-variables.pdf}.
In the case of FRB widths, one is concerned with the 
final distribution of observed pulse duration,

\begin{equation}
    n(t_{\mathrm{obs}}) = n_{t_{\mathrm{obs}}}(f(t_i)).
\end{equation}
\vspace{5pt}

\noindent The function $f$ is given by Eq.~\ref{eq-tobs}, 
and its inverse is, 

\begin{equation}
    t_i = f^{-1}(t_{\mathrm{obs}}) = \sqrt{\,t_{\mathrm{obs}}^2 - t_s^2 - t_{DM}^2}.
\end{equation}
\vspace{5pt}

The underlying FRB pulse width distribution, 
$n_{t_i}$, is transformed into an observed distribution as,

\begin{equation}
    n(t_{obs}) = n_{t_i}\left ( f^{-1}(t_{obs})\right )\frac{df^{-1}}{dt_{obs}},
    \label{eq-finv}
\end{equation}
\vspace{5pt}

\noindent Differentiating $f^{-1}$ and inserting it into Eq.~\ref{eq-finv}, 
one finds that the resultant 
observed width distribution at each DM is,

\begin{equation}
    n(t_{\mathrm{obs}}) = n_{t_i}\left ( \sqrt{\,t_{\mathrm{obs}}^2 - t_s^2 - t_{DM}^2} \right ) \frac{t_{\mathrm{obs}}}{\sqrt{\,t_{\mathrm{obs}}^2 - t_s^2 - t_{DM}^2}}.
\end{equation}
\vspace{5pt}

\noindent As an example, if the underlying width distribution is uniform, 
then $n_{t_i}(x)$ is constant and the observed pulse widths have 

\begin{equation}
    n(t_{obs}) = \frac{t_{obs}}{\sqrt{\,t_{obs}^2 - t_s^2 - t_{DM}^2}}.
\end{equation}
\vspace{5pt}

\noindent If the intrinsic width distribution is 
lognormal, one gets

\begin{equation}
    n(t_{obs}) = \mathrm{exp}\left \{ -\log^2\left(\sqrt{\,t_{obs}^2 - t_s^2 - t_{DM}^2}\right ) \right \} \frac{t_{obs}}{\sqrt{\,t_{obs}^2 - t_s^2 - t_{DM}^2}}.
\end{equation}

\begin{figure}
  \centering
    \includegraphics[width=0.48\textwidth]{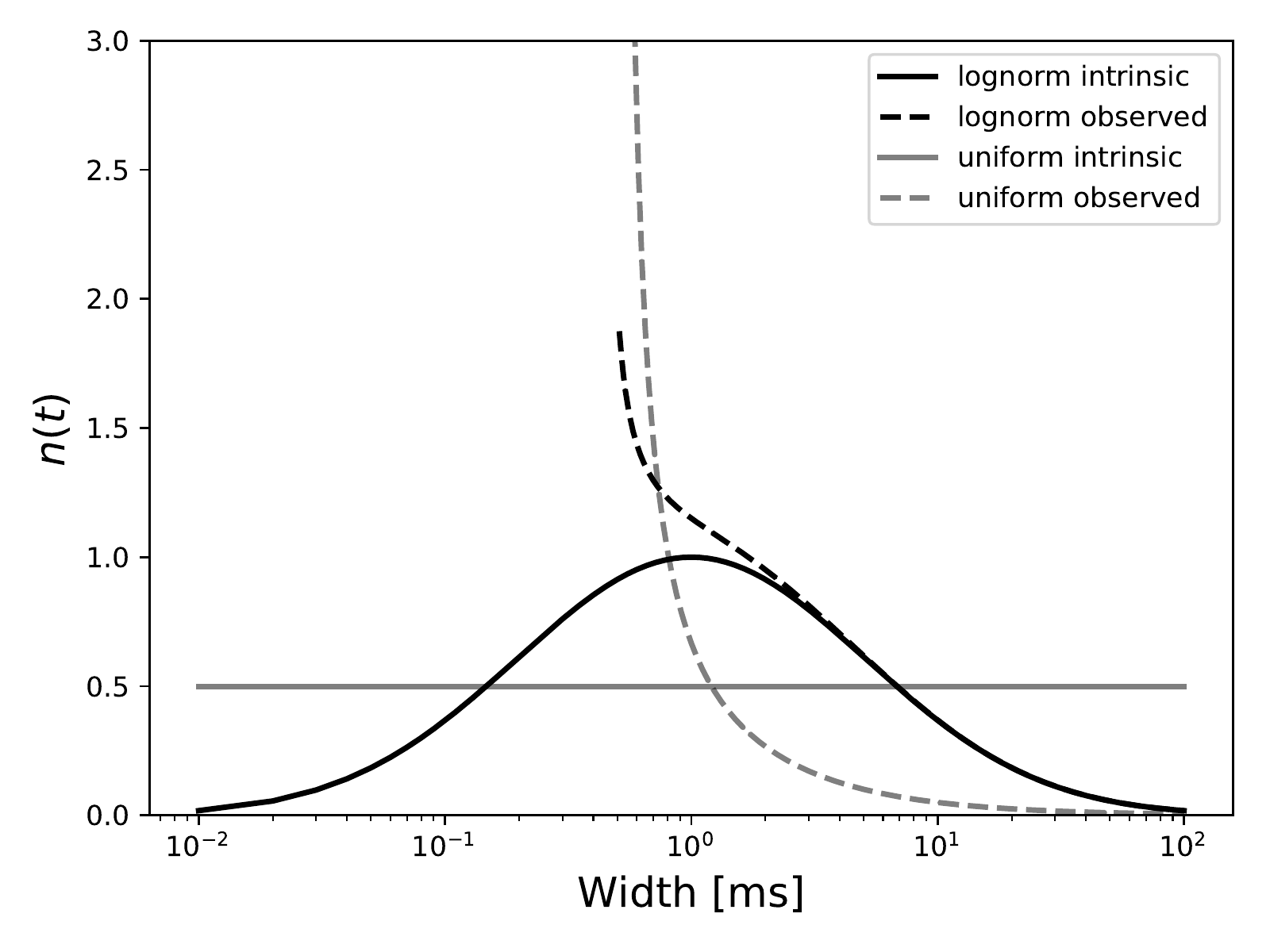}
  \caption{The underlying (solid) and observed (dashed) pulse 
  width distributions for a hypothetical survey with 0.5\,ms 
  sampling time. The black and grey curves represent a log-normal 
  and uniform input pdf, respectively. This figure 
  assumes dispersion smearing is small compared 
  to the sampling time.}
  \label{fig-width-obs}
\end{figure}

The intrinsic and observed width distributions are shown 
in solid and dashed curves, respectively, 
in Fig.~\ref{fig-width-obs}. The plot assumes 
DM smearing is significantly less than the sampling time, 
in this case, so no DM distribution was assumed. 
These functions are not defined 
for $t_{obs} \leq \sqrt{t_s^2 + t_{DM}^2}$ because 
one cannot detect an FRB that is narrower than the 
smearing timescale. That is why the observed PDF in 
Fig.~\ref{fig-width-obs} asymptotes to the instrumental 
smearing width, effectively limiting the 
detection width to the quadrature sum of dispersion 
smearing and sampling time. 

\subsection{Optimising survey resolution}
\label{sect-optimise}
\begin{figure*}
    \begin{center}
	\includegraphics[width=0.85\textwidth]{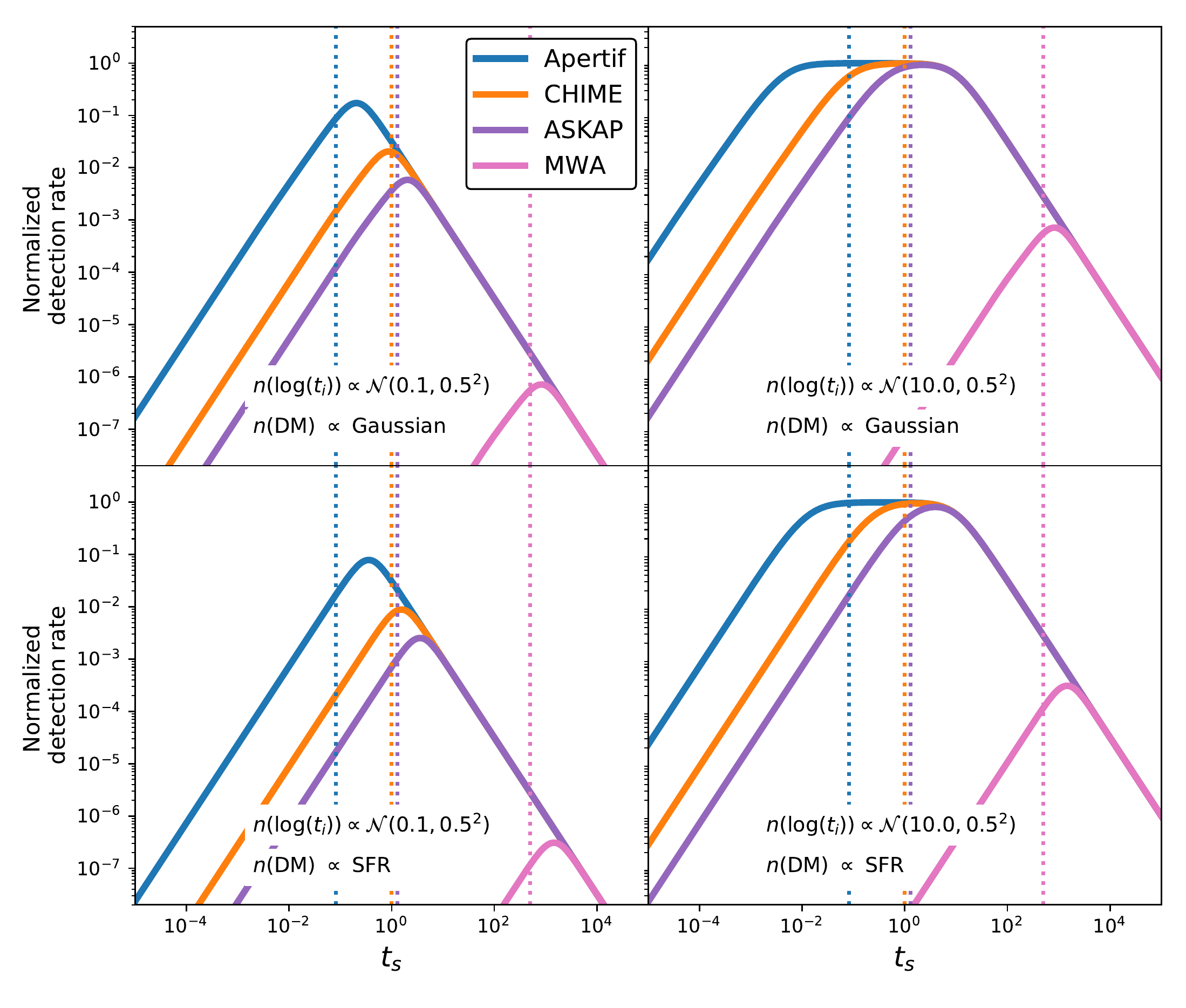}
    \caption{The FRB detection rate of several surveys normalised 
    to their detection rate if they had perfect time and frequency resolution, plotted
    as a function of sampling time. These are not absolute rates 
    to be compared between surveys, but a measure of their individual incompleteness due 
    to smearing. 
    I assume that each survey's temporal and spectral
    resolution are limited by data rates, meaning $t_s\Delta\nu$ is constant. 
    Dashed vertical lines indicate each telescope's nominal sampling time; the curve's peak 
    indicates an optimal sampling time for a given DM and intrinsic pulse width 
    distribution.}
    \label{fig-optimize}
    \end{center}
\end{figure*}

Modern digitisers sample the sky's electric field at an enormous rate, 
providing very fine temporal resolution at the telescope's front-end 
of roughly the inverse bandwidth. However, signal processing 
constraints mean data must be  
binned down in both time and frequency after the data 
are squared or correlated. Such throughput 
limitations often lead to the constraint that the 
product of the time and frequency resolution 
must be constant, 

\begin{equation}
    \Delta\nu \, t_s \equiv K\,\,\,[\mathrm{MHz\,ms}].
\end{equation}

\vspace{5pt}

\noindent In other words, if one wants twice as many 
frequency channels, one must lose a factor of
two in time resolution. 

Each survey will have its own value of $K$, and depending on the 
underlying DM, brightness, and pulse width distributions, 
a pair of ($t_s$, $\Delta\nu$) can be chosen to optimise 
FRB detection rate. This is done by differentiating 
$\mathcal{R}_{\hat{k}}$ in Eq.~\ref{eq-rate_sa} 
with respect to $t_s$ after replacing $\Delta\nu$ 
with $K/t_s$. In Fig.~\ref{fig-optimize} I plot the 
FRB detection rates for several surveys, assuming 
they are constrained by current spectral and temporal resolution, 
as a function of sampling time. As before, I have used simplifying assumptions about the intrinsic width and DM distributions. The two toy models for $n(\mathrm{DM})$, 
described in Sect.~\ref{sect-formalism}, assume standard 
candles and no $k$-corrections. I also assume no correlations 
between any input variables, such as brightness, 
DM, and intrinsic width.






Fig.~\ref{fig-optimize} calculates rates 
by summing over all DMs and intrinsic pulse widths 
to find the largest overall detection rate. 
However, depending on the survey's scientific interests, 
the goal may not be simply to maximize the number of detected 
events, but to maximize sensitivity to certain types of bursts.
For example, high-DM events come either from the greatest 
distances or the most dense environments, so surveys may want 
to upchannelise at the expense of time resolution.
A further 
constraint exists for telescopes that search formed beams in 
realtime, like Apertif and CHIME. If the 
conserved quantity in data rate is instead $t_s \Delta\nu\,N_{\mathrm{beam}}$, 
then the brightness distribution parameter $\alpha$ becomes more important. 
The trade-off between FoV and sensitivity is largely determined 
by source counts. If $\alpha$ turns out to be large and the brightness
distribution is steep, it may make sense 
to sacrifice number of beams for better 
temporal and spectral resolution. 

\subsection{Scattering}
\label{scattering}

Instead of attempting to model temporal scattering 
of FRBs explicitly, I have folded it into the intrinsic pulse 
width $t_i$. This is because this paper is concerned primarily with 
instrumental selection effects on DM, width, and brightness,
so astrophysical pulse broadening 
will simply alter the underlying width distribution of FRBs, $n(t_i)$.
Nonetheless, scattering is strongly frequency dependent ($\tau \propto \nu^{-4}$), meaning 
$n(t_i)$ will be a function of frequency. One should therefore expect 
the true widths of FRBs at low frequencies to be wider 
than at higher frequencies and surveys at different bands 
will suffer the deleterious effects of smearing 
differently. For this reason, when the same $n(t_i)$ is used
for surveys observing at different bands, as in
Fig.~\ref{fig-optimize}, it may be prudent to not compare the telescopes to 
one another directly, in case scattering significantly changes the 
distributions of $t_i$ as a function of frequency. As an example,
if CHIME continues to find that a significant fraction of FRBs 
between 400--800\,MHz are scattered, the optimisations 
described in this section may be less valuable there than at 1400\,MHz. To date, details about the
scattering statistics of FRBs are difficult to constrain.
CHIME found a majority of events \citep{chime2019a} significantly scattered (8/13), but ASKAP 
found evidence for scattering in just 3 of 20 events 
\citep{shannon2018}. Slightly less than 
half of Parkes FRBs are scattered \citep{petrofffrbcat, ravi2019}, but at this point we cannot know if such differences are due to scattering trends in frequency or DM, and which are due to instrumental selection effects.

\section{FRB Data}
\label{sect-data}

\subsection{Observed width distribution}
\begin{figure}
  \centering
    \includegraphics[width=0.46\textwidth]{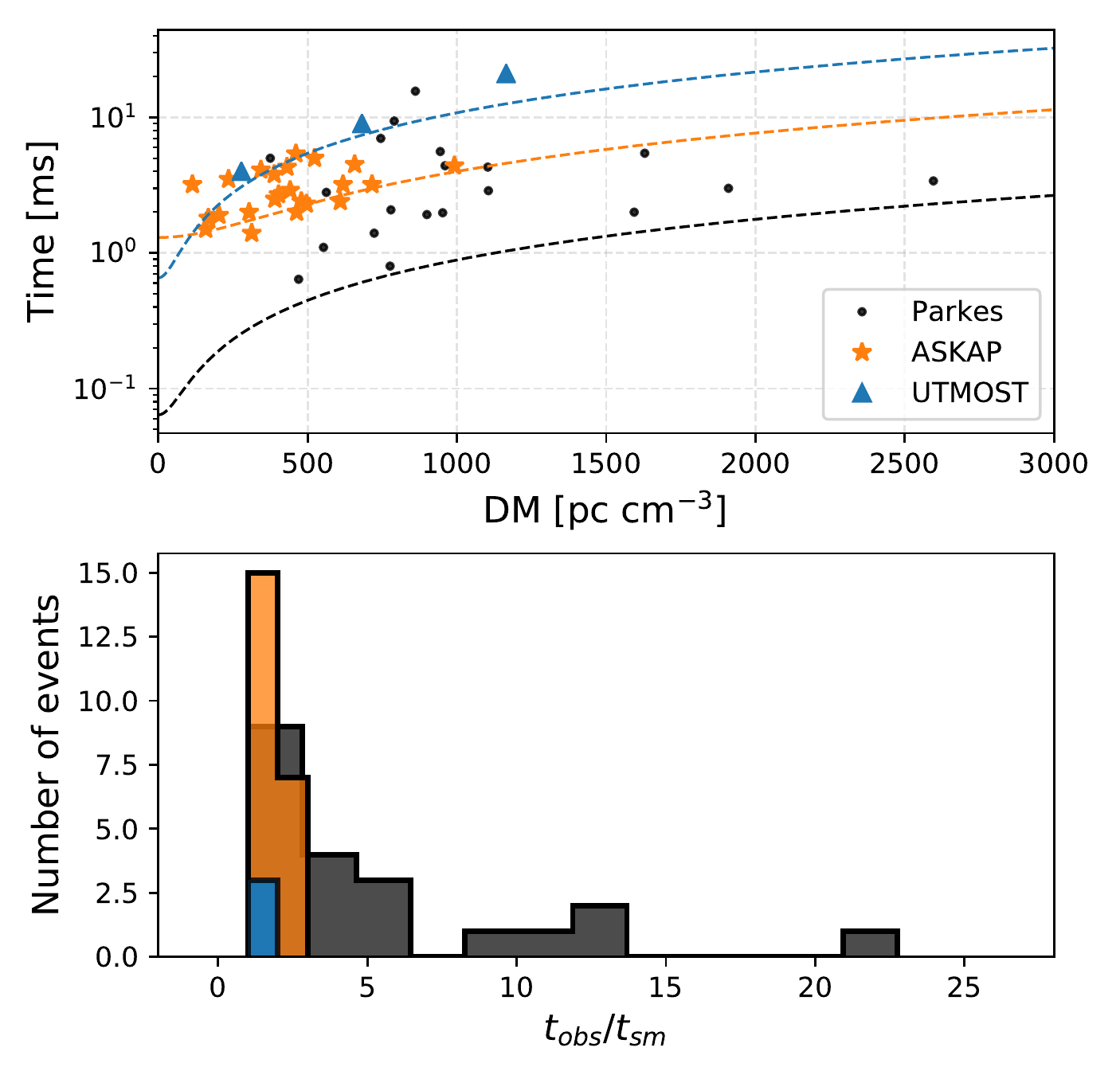}
  \vspace{-2pt}
  \caption{The observed pulse widths of 
    FRBs. The top panel shows width as a function of DM for 
    three surveys along with their theoretical smearing curves ($t_{sm}$). 
    The bottom panel shows the distributions of observed widths
    for the same three surveys, 
    normalised by the fundamental smearing limit at their 
    respective FRBs and with their back-ends. Nearly all 
    ASKAP and UTMOST FRBs `hug' the smearing curve.}
  \label{fig-1}
\end{figure}
To date, blind FRB detections have had widths between 
0.64--21\,ms, which is why FRBs are often described 
as millisecond-duration pulses. 
These detections were made with instruments 
whose smearing timescale is $\sim$\,millisecond at the 
relevant DMs, so little is known about the lower-limit 
of FRB widths. In Fig.~\ref{fig-1} I plot their observed 
widths as a function of DM as well as the distribution of 
their widths normalised by the smearing timescale, 
$t_{sm} = \sqrt{t^2_s + t^2_{\rm DM}}$.

The values $t_{\rm obs}$ plotted in Fig.~\ref{fig-1} 
are the detected pulse durations, typically
boxcar widths used by dedispersion codes blindly searching 
intensity data. The UTMOST FRBs are 
the telescope's first three detections, made before the system's time and frequency 
resolution were improved \citep{caleb2017}. The Parkes and ASKAP 
sources come from FRBCAT, with 21 and 23 FRBs respectively 
\citep{petrofffrbcat}. I do not include the 13 
pre-commissioning CHIME FRBs because their reported pulse duration
is not a detected boxcar width, but 
the fit from a Markov Chain Monte Carlo (MCMC) procedure; CHIME's 
large fractional bandwidth allows for the separation of 
frequency-dependent broadening effects like dispersion smearing and 
scattering.

The bursts hug closely their respective 
temporal smearing curves shown in the upper panel. 
The histogram in the bottom panel suggest that there are many FRBs narrower 
than these telescope's smearing timescale, and the observed distributions 
are qualitatively similar to the analytical curves in Fig.~\ref{fig-width-obs}.

Despite the millisecond widths of blind detections, 
it is known that FRBs have structure on much shorter timescales. 
This is because FRBs can be coherently dedispersed 
if voltage data are available, allowing for near-perfect 
time and frequency resolution. The first two sources 
to be coherently dedispersed offer insight into the temporal properties 
of FRBs. FRB121102 was the first source found to repeat, 
and its coherently dedispersed bursts have been as narrow as 
$\leq$\,30$\mu$s \citep{michilli2018}. FRB170827 was detected 
at UTMOST with incoherent dedispersion, but a voltage dump was triggered in 
real time. This allowed for coherent dedispersion that 
revealed microstructure of $\sim$\,30$\mu$s \citep{farah2018}. 
Similarly, \citet{ravi2016} argued that the 
spectral structure in FRB150807---which was unresolved even at 350\,$\mu$s---implied that its time profile  
was dominated by structure of a few microseconds.

It is worth noting that while these are the narrowest 
FRB components ever detected, the full burst 
often lasts longer than the microstructure.
Since they were found with incoherent back-ends 
with millisecond smearing timescales, both UTMOST and PALFA were biased to see FRBs whose duration aligned with their searchable boxcar widths. Until blind searches at 10s to 100s of microseconds are done, 
we will not know how many narrow FRBs exist.

\begin{figure*}
  \centering
    \includegraphics[width=0.7\textwidth]{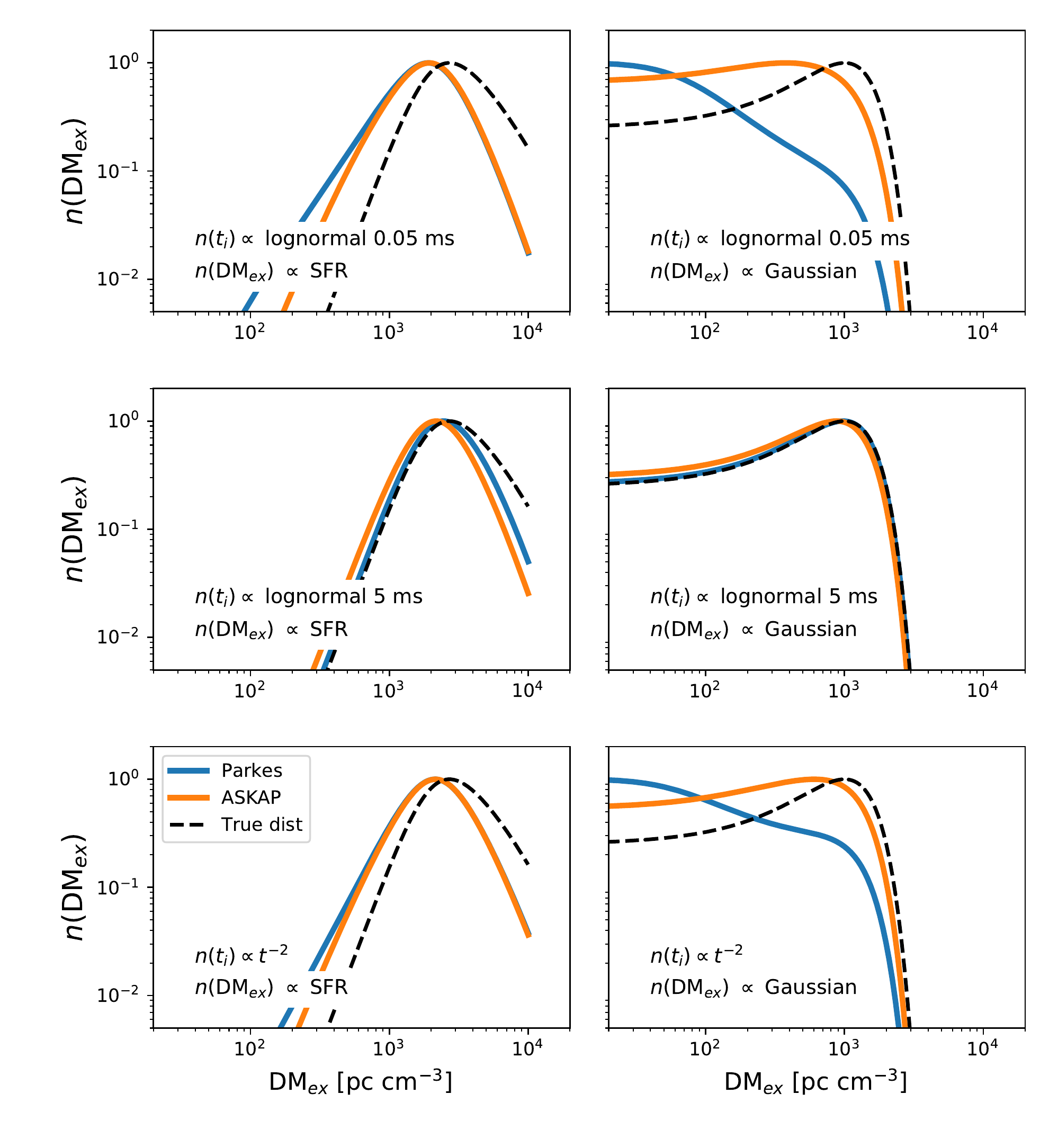}
  \vspace{-2pt}
  \caption{The simulated observed DM distribution for six combinations of 
          intrinsic FRB widths and DMs. Each panel shows 
          the distribution of detected DMs at Parkes (blue) and 
          ASKAP (orange) as well as the underyling DM distribution (black, dashed). Each is normalised to peak at 1. 
          The left column uses a true DM distribution that 
          tracks the star-formation rate, and the right column 
          is Gaussian with mean 1000\,pc\,cm$^{-3}$, 
          standard deviation 500\,pc\,cm$^{-3}$. The first, second, 
          and third rows use intrinsic width distributions that are
          lognormal with mean 0.05\,ms, lognormal with mean 5\,ms, 
          and loguniform respectively.}
  \label{fig-obs-dm}
\end{figure*}

\subsection{ASKAP and Parkes}
The observed extragalactic 
DM distribution differs significantly between 
Parkes FRBs and those detected with ASKAP, with lower DM$_{ex}$ 
events found with the latter. 
In fly's-eye mode, ASKAP is roughly 50 times less sensitive than 
Parkes and because FRB surveys have been 
flux-limited thus far, ASKAP necessarily detects 
much brighter FRBs than 
Parkes. Therefore, there may be evidence that 
extragalactic DM anti-correlates 
with brightness, which is expected 
for many luminosity functions in the cosmological FRB scenario \citep{shannon2018}.

However, as shown in Section~\ref{sec-dmdist} the observed 
DM distribution is transformed by the system's response, which 
are different at Parkes and ASKAP. The extent to which the 
observed DMs diverge will depend on the unknown \textit{true}
intrinsic pulse widths and DMs of the FRB population, 
and the time/frequency resolution and observing frequency of the instruments. Thus, 
one can ask the question, ``Which underlying width/DM$_{ex}$ distributions 
can produce such a discrepancy in the observed DMs that is 
purely due to instrumental smearing?''. 
In Fig.~\ref{fig-obs-dm} I plot the normalised
observed DM distributions at Parkes (blue) 
and ASKAP (orange) for six combinations of underlying width and 
DM$_{ex}$ inputs, assuming Euclidean source counts. 
The left column uses a true DM$_{ex}$ distribution (black, dashed curves) 
that traces the star-formation rate; the right column assumes a Gaussian 
intrinsic DM distribution, peaking at 1000\,pc\,cm$^{-3}$. 
Each of the three rows corresponds to a 
different underlying width distribution. 

I find that the the shape and scaling (detection rate) 
of the observed DM distribution can differ greatly from the 
true DM distribution. The observed distributions at Parkes 
and ASKAP can also vary significantly, due to their different time and 
frequency resolution (see the bottom right panel of Fig.~\ref{fig-obs-dm}). 
However, it is difficult to
shift the peaks of the ASKAP and Parkes DMs with instrumental smearing 
alone, unless contrived input widths and DM distributions are chosen. 
Therefore, 
I reach the same conclusion as \citet{shannon2018}, 
that the origin of the particular DM distribution discrepancy 
between Parkes and ASKAP is likely not due to 
smearing alone. 
My results do differ from their analysis regarding the behaviour 
of the distribution at high DMs. \citet{shannon2018}
write that any difference in the high-DM distribution at Parkes 
and ASKAP reflect intrinsic properties of the bursts detected. But 
in Fig.~\ref{fig-recovered} and Fig.~\ref{fig-obs-dm} it is clear 
that the observed DMs depend strongly on the chosen 
pulse width and DM distributions, which are not yet known. 
I also note that other selection effects, such as 
beams or search software, 
are not ruled out as the origin of the disparity in median observed DMs.



\section{Discussion}
\label{sec-discussion}
\subsection{Correlations between observables}
The formalism described in Sect.~\ref{sect-formalism} 
holds for arbitrary input distributions. 
Some of the figures, however, were generated assuming 
the relevant FRB observables are largely uncorrelated. 
Depending on what the nature of the FRB population turns out to be, 
this assumption may not be valid. Below I consider 
possible correlations between the three FRB variables 
focused on in this paper, namely brightness, DM, and pulse width.

\begin{figure}
\includegraphics[width=0.48\textwidth,trim={0.5cm 0cm 0cm 0cm}]{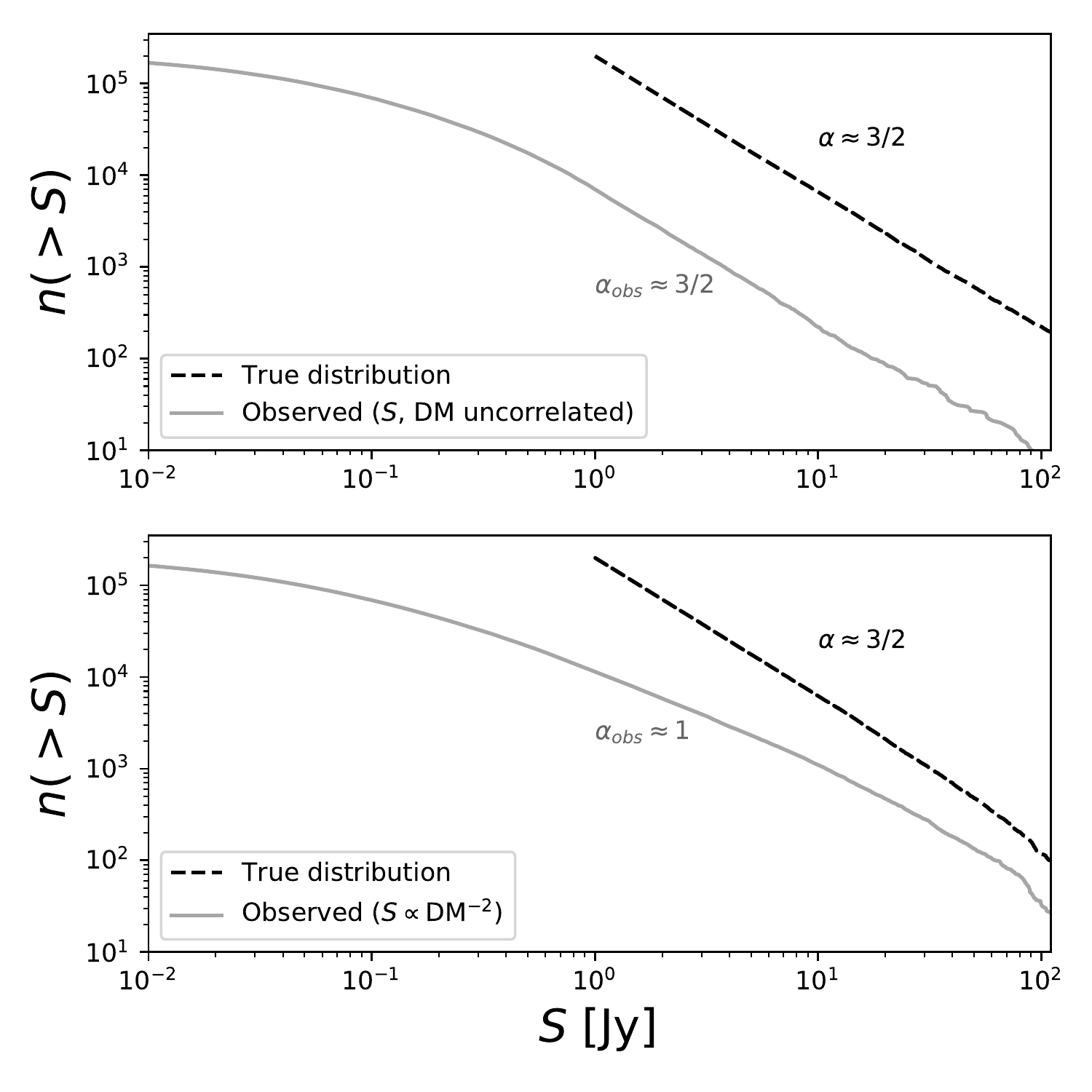}
\vspace{-10pt}
\caption{The effect of intra-channel dispersion smearing on 
           the observed source counts slope. Plotted are the 
           cumulative FRB brightness 
           distributions for two simulated datasets: 
           One in which DM and flux density are not correlated (top panel)
           and one in which DM is a distance measure, effecting an 
           inverse square relation between brightness and dispersion (bottom panel). 
           In both cases many events fall below the detection threshold due 
           to smearing, but in the latter case the apparent 
           source counts slope is significantly flattened due to 
           the correlation between brightness and DM.}
\label{fig-alpha}
\end{figure}

\subsubsection{DM / brightness and $\log N$--$\log S$}
There are a number of possible sources of covariance between 
DM and brightness. These include local phenomena, such 
as free-free absorption in a dense, dispersive plasma
environment; cosmological effects like the redshifting of the 
emission spectrum into the observing bandpass; and variation 
in the source population with redshift.

If DM and brightness are correlated, then the observed 
source counts of FRBs can be altered by intra-channel dispersion 
smearing. For most, but not all, plausible 
luminosity functions, a cosmological population of FRBs 
will show an anti-correlation between fluence and 
extragalactic DM because dispersion is a proxy 
for distance in those cases \citep{niino2018}; there may 
already be evidence for this in the data \citep{shannon2018}. 
If flux density or fluence do scale inversely with DM, the 
smearing effects described in this paper will result in an 
apparent flattening of the brightness distribution. This is 
because high-DM events, which are also the dimmest 
in that scenario, are more susceptible to dispersion smearing 
and a greater fraction of them will fall below the detection 
threshold. 
This results in a relative abundance of bright FRBs 
and smaller measured values of $\alpha$. I demonstrate 
the effect of smearing on source counts 
in the presence of a brightness/DM$_{ex}$ 
correlation in Fig.~\ref{fig-alpha}. 
In the simulation $2\times10^5$
FRBs are generated with a Euclidean 
flux distribution such that, $n(s)\propto s^{-3/2}$. 
Their DMs are Gaussian with mean 1000\,pc\,cm$^{-3}$ and standard deviation 500\,pc\,cm$^{-3}$ as before, and pulse widths are lognormal. 
Intra-channel dispersion smearing is 
dominant over sampling time. 
Brightness and DM are either uncorrelated (top panel) or 
perfectly correlated (bottom panel), with 
DM\,$\propto 1/\sqrt{s}$. In the correlated case, 
the logarithmic source counts 
are flattened and the measured power-law index is lower than 
the input Euclidean value. The effect is 
more pronounced for larger values of $\alpha$ since 
a steep brightness distribution means more events are 
clustered around the S/N threshold.


\subsubsection{DM / pulse width}
DM and \textit{observed} pulse width are trivially 
correlated, due to dispersion smearing.
Extragalactic DM and intrinsic pulse width (as they arrive at 
our receivers) may also be correlated. One source of this 
is cosmological time dilation if DM scales with distance, since 
\begin{equation}
    t_i = t_e\,(1+z) \propto\,\mathrm{DM}_{\mathrm{ex}}.
\end{equation}

Another potential source of this correlation is if there were a DM-scattering 
relationship analogous to that of Galactic pulsars \citep{bhat2004}. 
This could arise 
if FRBs typically lived in dense ionic environments, 
where local electrons contributed a significant fraction of the source's 
extragalactic dispersion. Such dense plasma environments are more likely to 
have a scattering screen along the FRB's line-of-sight. In 
the framework of this paper, scattering alters only $n(t_i)$ because
I consider the intrinsic pulse width to be the burst duration 
as it arrives at our receivers. Regardless of the cause, if 
DM$_{ex}$ and $t_i$ do indeed vary positively with 
one another due to cosmological or propagation broadening, 
higher-DM events will be more difficult to detect. 
This is because in those cases fluence is conserved, but detection S/N is not.

Similar to the proposed DM-brightness relation
in ASKAP/Parkes sources, there 
may also be weak evidence for the aforementioned 
DM-scattering relationship: 
7 of the 13 CHIME bursts found during pre-commissioning 
show temporal scattering and they report a weak correlation 
between scattering time and DM \citep{chime2019a}. \citet{ravi2019} found 
weak evidence for this at 1400\,MHz as well.
Since the IGM itself is not expected to produce 
detectable temporal scattering (see, e.g., \citet{luangoldreich2014}), 
the DM$_{\mathrm{ex}}$-scattering and 
DM$_{\mathrm{ex}}$-brightness trends may appear to be at odds, 
at least superficially.

If these two results 
gain statistical significance and are found not to 
be selection effects, there are three ways to reconcile them.
One explanation is that CHIME is probing a different 
population of FRBs at 600\,MHz than is 
being probed at 1400\,MHz. Alternatively, FRBs could be significantly 
dispersed by both the IGM and the local environment. 
Indeed, the one source whose local environment has been examined closely,
FRB121102, appears to have 
roughly half of its extragalactic dispersion in the 
IGM and half in its host galaxy \citep{tendulkar2017, michilli2018}.
A third explanation is that FRBs are dispersed by the 
IGM and scattered by the circumgalactic medium (CGM) of 
intervening galaxies \citep{vedantham2019}. 
Scattering timescale increases with distance 
to the screen, which is maximised when the screen is halfway between 
the observer and the source. Therefore, if the scattering of distant FRBs 
is dominated by screens in CGM halfway to the source, there ought to be a positive DM$_{\mathrm{ex}}$-scattering correlation. A counterexample 
may be FRB110523: \citet{masui-2015b} argued that its 
temporal scattering was due to a nearby screen, based on an 
angular broadening upper limit set by the fact that FRB110523 
scintillates in our Galaxy; if that source were first scattered 
in the CGM of an intervening galaxy, its angular broadening would 
be large enough that it would no longer be a point source 
when arriving in the Milky Way.

\subsubsection{Pulse width / brightness}
Brightness and pulse width could be inversely 
related if width increased with distance, whether 
due to cosmological time dilation or DM smearing. There could 
also exist a correlation between brightness and pulse width in 
the presence of scattering, which broadens the pulse and 
decreases peak flux density. Depending on the 
relationship between the pulse luminosity distribution and 
the pulse energy distribution, the correlation may also be intrinsic. 
As a trivial example, if the total \textit{energy} emitted in a pulse is 
roughly constant (called a ``standard battery'' by \citet{macquart2018_2}) rather than the total luminosity 
(``standard candle''), then pulse width and peak flux density 
would anti-correlate.

\subsection{Giant pulses} 
Galactic pulsars emit single pulses with widths that span 
$\sim$\,9 orders-of-magnitude, the shortest of which 
are giant pulses (GPs). By analogy, 
if a blind survey were to search
for GPs in our own Galaxy with current FRB back-ends, 
most would be missed due to smearing, as they typically have widths 
$<<$\,ms. The Crab, for example, 
whose GPs have the highest brightness temperature of 
anything in the Universe ($\sim$\,$10^{41}$\,K),  
emits pulses that are $\sim$\,microsecond at 1.4\,GHz. 
No lower limit has been placed on its minimum pulse width, but 
\citet{hankins2003} found Crab ``nanoshots'' of duration $<2$\,ns. 
Similarly, the 
distribution of B1937+21's GP widths was found to peak at
350\,ns \citep{mckee2018}.
If ASKAP
were to search for a 1\,$\mu$s Crab GP, its S/N would be 
reduced by a factor of 36 due to sampling smearing alone. 
Therefore, if FRBs also come from young 
neutron stars \citep{connor-2016a, cordes2016}, and are similar 
to giant pulses from pulsars,
the majority may be significantly narrower than 1\,ms.

\subsection{Low-frequency surveys}

Blind FRB searches at frequencies below 200\,MHz
have been limited by the volume of parameter space they 
could search, primarily due to instrumental smearing 
\citep{coenen2014, tingay2015, rowlinson2016}. As an example, for a real-time 
survey on LOFAR at 145\,MHz, \citet{karastergiou2015} used 
DM$_{max}=320$\,pc\,cm$^{-3}$ to avoid 
dispersion smearing. For many cosmological 
FRB scenarios, this could preclude detecting $\geq90\%$ of FRBs.
Even in the most optimistic case in Fig.~\ref{fig-optimize}, where there 
are few high-DM FRBs and they are all 
relatively broad, MWA's current back-end 
allows for just $1\%$ of events would be recovered. 
The most recent results from 
LOFAR's Fast Radio Transient Search (FRATS) was limited 
to DMs less than 500\,pc\,cm$^{-3}$ \citep{sander2019}. 
In this case,
they were not limited by 
intra-channel or sampling smearing (the 
time and frequency resolution were 0.5\,ms and 
12\,kHz respectively) but by the maximum 
dispersion delay across the band that could fit 
within their buffer boards. 

Upper limits obtained from non-detections 
must be multiplied by the reciprocal of the 
survey's completeness in width and DM, just as
sky-coverage and brightness incompleteness 
are accounted for when constraints are made 
using beam size and a minimum detectable flux 
density. 
Width and DM completeness are 
difficult to know precisely, because computing 
them requires knowing the underlying distributions 
of FRBs. But there exist similar 
problems with beam size and brightness, 
and given the extent of incompleteness due 
to instrumental smearing 
in many cases, previous constraints at low frequencies 
could be significantly weakened when considering 
smearing.

Upcoming surveys will observe simultaneously 
at multiple frequencies.
MWA has already successfully shadowed and recorded data 
at 185\,MHz during three of ASKAP's 1.4\,GHz detections
\citep{sokolowski2018}. Nothing was found, allowing them to 
constrain the spectral index of those bursts over a 
decade in frequency. 
However, their 1.28\,MHz frequency channels and 
500\,ms sampling time result in smearing, which would reduce a
millisecond burst with a moderate DM by a factor of $\sim$30 in S/N. 
The Apertif LOFAR Exploration of the 
Radio Transient Sky (ALERT) will detect FRBs 
with Apertif at 1.4\,GHz and
trigger LOFAR at 150\,MHz. This system has the advantage 
of recording voltage data, and will save 5\,s 
of data in each frequency sub-band of LOFAR in order to 
extract the pulse's dispersion sweep. For this reason,
ALERT will be able to probe low-frequency 
FRB emission in an unprecedented way. The 
existence of temporally unresolved, unscattered 
FRBs detected down 
to the bottom of CHIME's observing band at 400\,MHz 
is promising for such programs \citep{chime2019a}.

\section{Conclusions}

I have introduced a formalism for calculating 
FRB detection rates, and I have incorporated
relevant burst and instrumental parameters that have previously 
been neglected. 
I have also presented a method for interpreting 
the distributions of FRB observables, focusing on DM, 
pulse width, and brightness. 
Depending on the intrinsic DM 
and pulse width distributions of FRBs, otherwise-similar telescopes 
can have orders-of-magnitude differences in detection rate depending 
on their time and frequency resolution, 
due only to the deleterious effects of temporal smearing 
(see, e.g., Fig~\ref{fig-rateti}). 
Perhaps more significantly, the interaction between 
the intrinsic distributions of FRB properties and 
the instruments that detect them can produce 
\textit{observed} distributions that look very 
different from the inputs (Fig~\ref{fig-width-obs}, Fig~\ref{fig-obs-dm}). 
I therefore caution against 
inferring physical properties about FRBs (luminosity function, 
spatial distribution, etc.) based on their observables 
without accounting for such selection effects, even when the 
analysis is done within a single survey.

I have argued that if there are substantial numbers of 
narrow FRBs, they would 
fall below our detection thresholds due to smearing. 
Based on the observed 
distributions of burst duration and the fact that coherently-dedispersed 
FRBs tend to show temporal structure at $<<$\,ms, a population 
of microsecond-duration events may be likely. 
Another piece of evidence for the existence of narrow bursts 
comes from the 13 pre-commissioning CHIME FRBs \citep{chime2019a}. By effectively 
fitting out frequency-dependent broadening effects like scattering 
and dispersion smearing, they found 4 events with duration 
$<500$\,$\mu$s, even though that is well below their 
smearing timescale. Those events were detected despite attenuation 
due to smearing, but there are likely less bright events of similar widths 
did fall below the S/N threshold. 
The same arguments could be made about FRB150807, 
which was temporally unresolved at 350\,$\mu$s and whose frequency spectrum 
implied structure at the level of several microseconds \citep{ravi2016}.

To test the hypotheses that narrow or high-DM bursts are being missed, 
a Kolmogorov-Smirnov test 
could be run on current data from different telescopes, 
comparing the distribution of $t_{obs}/t_{sm}$ with those produced 
by simulated DM and width distributions.
Another approach would be to alter the balance 
between time and frequency resolution on throughput-limited surveys
in order to increase sensitivity in specific regions 
of width/DM space. For example, if FRBs are to be searched 
for at DM\,$>\,3000$\,pc\,cm$^{-3}$, a survey could sacrifice time resolution 
or number of beams for increased channelization. 
A single telescope could carry out two distinct surveys with different 
time and frequency resolutions in order to investigate the underlying distributions 
of width and DM.  
To that end, I have presented a method for 
calculating the rate-optimized time and frequency resolution 
at a given survey was presented in Sect.~\ref{sect-optimise}. 

Finally, I considered intrinsic correlations between various 
FRB parameters. I showed how dispersion smearing can lead to an 
apparent flattening of FRB source counts in the case where 
DM and brightness are inversely related. I also discussed 
the potential physical and instrumental causes for a 
positive correlation between DM and detected pulse width. 
As we enter an era in which multiple surveys will each 
have considerable collections of FRBs, the community will need end-to-end 
simulation pipelines that include instrumental selection 
effects and population synthesis 
(e.g.  {\tt frbpoppy}\footnote{https://github.com/davidgardenier/frbpoppy} 
Gardenier et al. in prep (2019)). Still, due to the subtle 
selection effects involved in searching for and interpreting 
FRB detections, it is vital to understand such phenomena from the 
ground up.

\section*{Acknowledgements}

I thank Emily Petroff for helpful notes 
on the manuscript, as well as the anonymous referee 
for valuable suggestions. I also thank the organisers
of FRB2019 Amsterdam, which provided valuable discussion related
to this work. 
I receive funding from the European Research 
Council under the European Union's Seventh 
Framework Programme (FP/2007-2013) / ERC Grant Agreement n. 617199.





\bibliography{widthpaper}

\begin{thebibliography}{}
\makeatletter
\relax
\def\mn@urlcharsother{\let\do\@makeother \do\$\do\&\do\#\do\^\do\_\do\%\do\~}
\def\mn@doi{\begingroup\mn@urlcharsother \@ifnextchar [ {\mn@doi@}
  {\mn@doi@[]}}
\def\mn@doi@[#1]#2{\def\@tempa{#1}\ifx\@tempa\@empty \href
  {http://dx.doi.org/#2} {doi:#2}\else \href {http://dx.doi.org/#2} {#1}\fi
  \endgroup}
\def\mn@eprint#1#2{\mn@eprint@#1:#2::\@nil}
\def\mn@eprint@arXiv#1{\href {http://arxiv.org/abs/#1} {{\tt arXiv:#1}}}
\def\mn@eprint@dblp#1{\href {http://dblp.uni-trier.de/rec/bibtex/#1.xml}
  {dblp:#1}}
\def\mn@eprint@#1:#2:#3:#4\@nil{\def\@tempa {#1}\def\@tempb {#2}\def\@tempc
  {#3}\ifx \@tempc \@empty \let \@tempc \@tempb \let \@tempb \@tempa \fi \ifx
  \@tempb \@empty \def\@tempb {arXiv}\fi \@ifundefined
  {mn@eprint@\@tempb}{\@tempb:\@tempc}{\expandafter \expandafter \csname
  mn@eprint@\@tempb\endcsname \expandafter{\@tempc}}}

\bibitem[\protect\citeauthoryear{{Amiri} et~al.,}{{Amiri}
  et~al.}{2017}]{connor2017}
{Amiri} M.,  et~al., 2017, \mn@doi [\apj] {10.3847/1538-4357/aa713f}, \href
  {http://adsabs.harvard.edu/abs/2017ApJ...844..161A} {844, 161}

\bibitem[\protect\citeauthoryear{{Bailes} et~al.,}{{Bailes}
  et~al.}{2017}]{utmost2017}
{Bailes} M.,  et~al., 2017, \mn@doi [\pasa] {10.1017/pasa.2017.39}, \href
  {http://adsabs.harvard.edu/abs/2017PASA...34...45B} {34, e045}

\bibitem[\protect\citeauthoryear{{Bannister} et~al.,}{{Bannister}
  et~al.}{2017}]{bannister2017}
{Bannister} K.~W.,  et~al., 2017, \mn@doi [\apjl] {10.3847/2041-8213/aa71ff},
  \href {http://adsabs.harvard.edu/abs/2017ApJ...841L..12B} {841, L12}

\bibitem[\protect\citeauthoryear{{Bhat}, {Cordes}, {Camilo}, {Nice}  \&
  {Lorimer}}{{Bhat} et~al.}{2004}]{bhat2004}
{Bhat} N.~D.~R.,  {Cordes} J.~M.,  {Camilo} F.,  {Nice} D.~J.,   {Lorimer}
  D.~R.,  2004, \mn@doi [\apj] {10.1086/382680}, \href
  {http://adsabs.harvard.edu/abs/2004ApJ...605..759B} {605, 759}

\bibitem[\protect\citeauthoryear{{CHIME/FRB Collaboration} et~al.,}{{CHIME/FRB
  Collaboration} et~al.}{2018}]{chime2018overview}
{CHIME/FRB Collaboration} et~al., 2018, \mn@doi [\apj]
  {10.3847/1538-4357/aad188}, \href
  {http://adsabs.harvard.edu/abs/2018ApJ...863...48C} {863, 48}

\bibitem[\protect\citeauthoryear{{CHIME/FRB Collaboration} et~al.,}{{CHIME/FRB
  Collaboration} et~al.}{2019a}]{chime2019a}
{CHIME/FRB Collaboration} et~al., 2019a, \mn@doi [\nat]
  {10.1038/s41586-018-0867-7}, \href
  {http://adsabs.harvard.edu/abs/2019Natur.566..230C} {566, 230}

\bibitem[\protect\citeauthoryear{{CHIME/FRB Collaboration} et~al.,}{{CHIME/FRB
  Collaboration} et~al.}{2019b}]{chime2019r2}
{CHIME/FRB Collaboration} et~al., 2019b, \mn@doi [\nat]
  {10.1038/s41586-018-0864-x}, \href
  {http://adsabs.harvard.edu/abs/2019Natur.566..235C} {566, 235}

\bibitem[\protect\citeauthoryear{{Caleb}, {Flynn}, {Bailes}, {Barr},
  {Hunstead}, {Keane}, {Ravi}  \& {van Straten}}{{Caleb}
  et~al.}{2016}]{caleb2016a}
{Caleb} M.,  {Flynn} C.,  {Bailes} M.,  {Barr} E.~D.,  {Hunstead} R.~W.,
  {Keane} E.~F.,  {Ravi} V.,   {van Straten} W.,  2016, \mn@doi [\mnras]
  {10.1093/mnras/stw175}, \href
  {http://adsabs.harvard.edu/abs/2016MNRAS.458..708C} {458, 708}

\bibitem[\protect\citeauthoryear{{Caleb} et~al.,}{{Caleb}
  et~al.}{2017}]{caleb2017}
{Caleb} M.,  et~al., 2017, \mn@doi [\mnras] {10.1093/mnras/stx638}, \href
  {http://adsabs.harvard.edu/abs/2017MNRAS.468.3746C} {468, 3746}

\bibitem[\protect\citeauthoryear{{Coenen} et~al.,}{{Coenen}
  et~al.}{2014}]{coenen2014}
{Coenen} T.,  et~al., 2014, \mn@doi [\aap] {10.1051/0004-6361/201424495}, \href
  {http://adsabs.harvard.edu/abs/2014A%26A...570A..60C} {570, A60}

\bibitem[\protect\citeauthoryear{{Connor}, {Sievers}  \& {Pen}}{{Connor}
  et~al.}{2016a}]{connor-2016a}
{Connor} L.,  {Sievers} J.,   {Pen} U.-L.,  2016a, \mn@doi [\mnras]
  {10.1093/mnrasl/slv124}, \href
  {http://adsabs.harvard.edu/abs/2016MNRAS.458L..19C} {458, L19}

\bibitem[\protect\citeauthoryear{{Connor}, {Lin}, {Masui}, {Oppermann}, {Pen},
  {Peterson}, {Roman}  \& {Sievers}}{{Connor} et~al.}{2016b}]{connor2016b}
{Connor} L.,  {Lin} H.-H.,  {Masui} K.,  {Oppermann} N.,  {Pen} U.-L.,
  {Peterson} J.~B.,  {Roman} A.,   {Sievers} J.,  2016b, \mn@doi [\mnras]
  {10.1093/mnras/stw907}, \href
  {http://adsabs.harvard.edu/abs/2016MNRAS.460.1054C} {460, 1054}

\bibitem[\protect\citeauthoryear{{Cordes} \& {McLaughlin}}{{Cordes} \&
  {McLaughlin}}{2003}]{cordes2003}
{Cordes} J.~M.,  {McLaughlin} M.~A.,  2003, \mn@doi [\apj] {10.1086/378231},
  \href {http://adsabs.harvard.edu/abs/2003ApJ...596.1142C} {596, 1142}

\bibitem[\protect\citeauthoryear{{Cordes} \& {Wasserman}}{{Cordes} \&
  {Wasserman}}{2016}]{cordes2016}
{Cordes} J.~M.,  {Wasserman} I.,  2016, \mn@doi [\mnras]
  {10.1093/mnras/stv2948}, \href
  {http://adsabs.harvard.edu/abs/2016MNRAS.457..232C} {457, 232}

\bibitem[\protect\citeauthoryear{{Farah} et~al.,}{{Farah}
  et~al.}{2018}]{farah2018}
{Farah} W.,  et~al., 2018, \mn@doi [\mnras] {10.1093/mnras/sty1122}, \href
  {http://adsabs.harvard.edu/abs/2018MNRAS.478.1209F} {478, 1209}

\bibitem[\protect\citeauthoryear{{Gajjar} et~al.,}{{Gajjar}
  et~al.}{2018}]{Gajjar2018}
{Gajjar} V.,  et~al., 2018, \mn@doi [\apj] {10.3847/1538-4357/aad005}, \href
  {http://adsabs.harvard.edu/abs/2018ApJ...863....2G} {863, 2}

\bibitem[\protect\citeauthoryear{{Hankins}, {Kern}, {Weatherall}  \&
  {Eilek}}{{Hankins} et~al.}{2003}]{hankins2003}
{Hankins} T.~H.,  {Kern} J.~S.,  {Weatherall} J.~C.,   {Eilek} J.~A.,  2003,
  \mn@doi [\nat] {10.1038/nature01477}, \href
  {http://adsabs.harvard.edu/abs/2003Natur.422..141H} {422, 141}

\bibitem[\protect\citeauthoryear{{James} et~al.,}{{James}
  et~al.}{2019}]{james2019b}
{James} C.~W.,  et~al., 2019, \mn@doi [\pasa] {10.1017/pasa.2019.1}, \href
  {http://adsabs.harvard.edu/abs/2019PASA...36....9J} {36, e009}

\bibitem[\protect\citeauthoryear{{Karastergiou} et~al.,}{{Karastergiou}
  et~al.}{2015}]{karastergiou2015}
{Karastergiou} A.,  et~al., 2015, \mn@doi [\mnras] {10.1093/mnras/stv1306},
  \href {http://adsabs.harvard.edu/abs/2015MNRAS.452.1254K} {452, 1254}

\bibitem[\protect\citeauthoryear{{Keane} \& {Petroff}}{{Keane} \&
  {Petroff}}{2015}]{keane2015}
{Keane} E.~F.,  {Petroff} E.,  2015, \mn@doi [\mnras] {10.1093/mnras/stu2650},
  \href {http://adsabs.harvard.edu/abs/2015MNRAS.447.2852K} {447, 2852}

\bibitem[\protect\citeauthoryear{{Lawrence}, {Vander Wiel}, {Law}, {Burke
  Spolaor}  \& {Bower}}{{Lawrence} et~al.}{2017}]{Lawrence2017}
{Lawrence} E.,  {Vander Wiel} S.,  {Law} C.,  {Burke Spolaor} S.,   {Bower}
  G.~C.,  2017, \mn@doi [\aj] {10.3847/1538-3881/aa844e}, \href
  {http://adsabs.harvard.edu/abs/2017AJ....154..117L} {154, 117}

\bibitem[\protect\citeauthoryear{{Lorimer}, {Bailes}, {McLaughlin}, {Narkevic}
  \& {Crawford}}{{Lorimer} et~al.}{2007}]{lorimer07}
{Lorimer} D.~R.,  {Bailes} M.,  {McLaughlin} M.~A.,  {Narkevic} D.~J.,
  {Crawford} F.,  2007, \mn@doi [Science] {10.1126/science.1147532}, \href
  {http://adsabs.harvard.edu/abs/2007Sci...318..777L} {318, 777}

\bibitem[\protect\citeauthoryear{{Luan} \& {Goldreich}}{{Luan} \&
  {Goldreich}}{2014}]{luangoldreich2014}
{Luan} J.,  {Goldreich} P.,  2014, \mn@doi [\apjl]
  {10.1088/2041-8205/785/2/L26}, \href
  {http://adsabs.harvard.edu/abs/2014ApJ...785L..26L} {785, L26}

\bibitem[\protect\citeauthoryear{{Macquart} \& {Ekers}}{{Macquart} \&
  {Ekers}}{2018a}]{macquart2018_1}
{Macquart} J.-P.,  {Ekers} R.~D.,  2018a, \mn@doi [\mnras]
  {10.1093/mnras/stx2825}, \href
  {http://adsabs.harvard.edu/abs/2018MNRAS.474.1900M} {474, 1900}

\bibitem[\protect\citeauthoryear{{Macquart} \& {Ekers}}{{Macquart} \&
  {Ekers}}{2018b}]{macquart2018_2}
{Macquart} J.-P.,  {Ekers} R.,  2018b, \mn@doi [\mnras]
  {10.1093/mnras/sty2083}, \href
  {http://adsabs.harvard.edu/abs/2018MNRAS.480.4211M} {480, 4211}

\bibitem[\protect\citeauthoryear{{Madau} \& {Dickinson}}{{Madau} \&
  {Dickinson}}{2014}]{madau2014}
{Madau} P.,  {Dickinson} M.,  2014, \mn@doi [\araa]
  {10.1146/annurev-astro-081811-125615}, \href
  {http://adsabs.harvard.edu/abs/2014ARA%26A..52..415M} {52, 415}

\bibitem[\protect\citeauthoryear{{Madhavacheril}, {Battaglia}, {Smith}  \&
  {Sievers}}{{Madhavacheril} et~al.}{2019}]{Madhavacheril2019}
{Madhavacheril} M.~S.,  {Battaglia} N.,  {Smith} K.~M.,   {Sievers} J.~L.,
  2019, arXiv e-prints, \href
  {http://adsabs.harvard.edu/abs/2019arXiv190102418M} {}

\bibitem[\protect\citeauthoryear{{Masui} \& {Sigurdson}}{{Masui} \&
  {Sigurdson}}{2015}]{masui2015a}
{Masui} K.~W.,  {Sigurdson} K.,  2015, \mn@doi [Physical Review Letters]
  {10.1103/PhysRevLett.115.121301}, \href
  {http://adsabs.harvard.edu/abs/2015PhRvL.115l1301M} {115, 121301}

\bibitem[\protect\citeauthoryear{{Masui} et~al.,}{{Masui}
  et~al.}{2015}]{masui-2015b}
{Masui} K.,  et~al., 2015, \mn@doi [\nat] {10.1038/nature15769}, \href
  {http://adsabs.harvard.edu/abs/2015Natur.528..523M} {528, 523}

\bibitem[\protect\citeauthoryear{{McKee} et~al.,}{{McKee}
  et~al.}{2018}]{mckee2018}
{McKee} J.~W.,  et~al., 2018, \mn@doi [\mnras] {10.1093/mnras/sty3058}, \href
  {http://adsabs.harvard.edu/abs/2018MNRAS.tmp.2917M} {}

\bibitem[\protect\citeauthoryear{{McQuinn}}{{McQuinn}}{2014}]{mcquinn2014}
{McQuinn} M.,  2014, \mn@doi [\apjl] {10.1088/2041-8205/780/2/L33}, \href
  {http://adsabs.harvard.edu/abs/2014ApJ...780L..33M} {780, L33}

\bibitem[\protect\citeauthoryear{{Michilli} et~al.,}{{Michilli}
  et~al.}{2018}]{michilli2018}
{Michilli} D.,  et~al., 2018, \mn@doi [\nat] {10.1038/nature25149}, \href
  {http://adsabs.harvard.edu/abs/2018Natur.553..182M} {553, 182}

\bibitem[\protect\citeauthoryear{{Niino}}{{Niino}}{2018}]{niino2018}
{Niino} Y.,  2018, \mn@doi [\apj] {10.3847/1538-4357/aab9a9}, \href
  {http://adsabs.harvard.edu/abs/2018ApJ...858....4N} {858, 4}

\bibitem[\protect\citeauthoryear{{Oppermann}, {Connor}  \& {Pen}}{{Oppermann}
  et~al.}{2016}]{Oppermann16}
{Oppermann} N.,  {Connor} L.~D.,   {Pen} U.-L.,  2016, \mn@doi [\mnras]
  {10.1093/mnras/stw1401}, \href
  {http://adsabs.harvard.edu/abs/2016MNRAS.461..984O} {461, 984}

\bibitem[\protect\citeauthoryear{{Petroff} et~al.,}{{Petroff}
  et~al.}{2016}]{petrofffrbcat}
{Petroff} E.,  et~al., 2016, \mn@doi [\pasa] {10.1017/pasa.2016.35}, \href
  {http://adsabs.harvard.edu/abs/2016PASA...33...45P} {33, e045}

\bibitem[\protect\citeauthoryear{{Ravi}}{{Ravi}}{2019}]{ravi2019}
{Ravi} V.,  2019, \mn@doi [\mnras] {10.1093/mnras/sty1551}, \href
  {http://adsabs.harvard.edu/abs/2019MNRAS.482.1966R} {482, 1966}

\bibitem[\protect\citeauthoryear{{Ravi} et~al.,}{{Ravi}
  et~al.}{2016}]{ravi2016}
{Ravi} V.,  et~al., 2016, \mn@doi [Science] {10.1126/science.aaf6807}, \href
  {http://adsabs.harvard.edu/abs/2016Sci...354.1249R} {354, 1249}

\bibitem[\protect\citeauthoryear{{Rowlinson} et~al.,}{{Rowlinson}
  et~al.}{2016}]{rowlinson2016}
{Rowlinson} A.,  et~al., 2016, \mn@doi [\mnras] {10.1093/mnras/stw451}, \href
  {http://adsabs.harvard.edu/abs/2016MNRAS.458.3506R} {458, 3506}

\bibitem[\protect\citeauthoryear{Sanidas, Caleb, Driessen, Morello, Rajwade  \&
  Stappers}{Sanidas et~al.}{2017}]{sanidas2018}
Sanidas S.,  Caleb M.,  Driessen L.,  Morello V.,  Rajwade K.,   Stappers
  B.~W.,  2017, \mn@doi [Proceedings of the International Astronomical Union]
  {10.1017/S1743921317009310}, 13, 406–407

\bibitem[\protect\citeauthoryear{{Shannon} et~al.,}{{Shannon}
  et~al.}{2018}]{shannon2018}
{Shannon} R.~M.,  et~al., 2018, \mn@doi [\nat] {10.1038/s41586-018-0588-y},
  \href {http://adsabs.harvard.edu/abs/2018Natur.562..386S} {562, 386}

\bibitem[\protect\citeauthoryear{{Sokolowski} et~al.,}{{Sokolowski}
  et~al.}{2018}]{sokolowski2018}
{Sokolowski} M.,  et~al., 2018, \mn@doi [\apjl] {10.3847/2041-8213/aae58d},
  \href {http://adsabs.harvard.edu/abs/2018ApJ...867L..12S} {867, L12}

\bibitem[\protect\citeauthoryear{{Spitler} et~al.,}{{Spitler}
  et~al.}{2014}]{spitler2014}
{Spitler} L.~G.,  et~al., 2014, \mn@doi [\apj] {10.1088/0004-637X/790/2/101},
  \href {http://adsabs.harvard.edu/abs/2014ApJ...790..101S} {790, 101}

\bibitem[\protect\citeauthoryear{{Spitler} et~al.,}{{Spitler}
  et~al.}{2016}]{spitler2016}
{Spitler} L.~G.,  et~al., 2016, \mn@doi [\nat] {10.1038/nature17168}, \href
  {http://adsabs.harvard.edu/abs/2016Natur.531..202S} {531, 202}

\bibitem[\protect\citeauthoryear{{Tendulkar} et~al.,}{{Tendulkar}
  et~al.}{2017}]{tendulkar2017}
{Tendulkar} S.~P.,  et~al., 2017, \mn@doi [\apjl] {10.3847/2041-8213/834/2/L7},
  \href {http://adsabs.harvard.edu/abs/2017ApJ...834L...7T} {834, L7}

\bibitem[\protect\citeauthoryear{{Thornton} et~al.,}{{Thornton}
  et~al.}{2013}]{thornton-2013}
{Thornton} D.,  et~al., 2013, \mn@doi [Science] {10.1126/science.1236789},
  \href {http://adsabs.harvard.edu/abs/2013Sci...341...53T} {341, 53}

\bibitem[\protect\citeauthoryear{{Tingay} et~al.,}{{Tingay}
  et~al.}{2015}]{tingay2015}
{Tingay} S.~J.,  et~al., 2015, \mn@doi [\aj] {10.1088/0004-6256/150/6/199},
  \href {http://adsabs.harvard.edu/abs/2015AJ....150..199T} {150, 199}

\bibitem[\protect\citeauthoryear{{Vedantham} \& {Phinney}}{{Vedantham} \&
  {Phinney}}{2019}]{vedantham2019}
{Vedantham} H.~K.,  {Phinney} E.~S.,  2019, \mn@doi [\mnras]
  {10.1093/mnras/sty2948}, \href
  {http://adsabs.harvard.edu/abs/2019MNRAS.483..971V} {483, 971}

\bibitem[\protect\citeauthoryear{{Vedantham}, {Ravi}, {Hallinan}  \&
  {Shannon}}{{Vedantham} et~al.}{2016}]{vedantham2016}
{Vedantham} H.~K.,  {Ravi} V.,  {Hallinan} G.,   {Shannon} R.~M.,  2016,
  \mn@doi [\apj] {10.3847/0004-637X/830/2/75}, \href
  {http://adsabs.harvard.edu/abs/2016ApJ...830...75V} {830, 75}

\bibitem[\protect\citeauthoryear{{ter Veen} et~al.,}{{ter Veen}
  et~al.}{2019}]{sander2019}
{ter Veen} S.,  et~al., 2019, \mn@doi [\aap] {10.1051/0004-6361/201732515},
  \href {http://adsabs.harvard.edu/abs/2019A%26A...621A..57T} {621, A57}

\bibitem[\protect\citeauthoryear{{van Leeuwen}}{{van Leeuwen}}{2014}]{leeu14}
{van Leeuwen} J.,  2014, in {Wozniak} P.~R.,  {Graham} M.~J.,  {Mahabal} A.~A.,
    {Seaman} R.,  eds, The Third Hot-wiring the Transient Universe Workshop. pp
  79--79

\makeatother
\end{thebibliography}
\bibliographystyle{mnras}





\label{lastpage}
\end{document}